\documentclass{aa}  
\usepackage{graphicx}
\usepackage{natbib,graphicx}
\bibpunct{(}{)}{;}{a}{}{,} 
\begin{document}
\title{A large $^{12}$C/$^{13}$C isotopic ratio in M~82 and NGC~253}

\author{
  Sergio Mart\'in\inst{\ref{inst1}}
  \and R. Aladro\inst{\ref{inst2}}
  \and J. Mart\'in-Pintado\inst{\ref{inst3}}
  \and R. Mauersberger \inst{\ref{inst4}}
}
\institute{
  European Southern Observatory, Alonso de C\'ordova 3107, Vitacura, Casilla 19001, Santiago 19, Chile\\
  \email{smartin@eso.org}\label{inst1}
  \and
  Instituto de Radioastronom\'ia Milim\'etrica (IRAM), Avda. Divina Pastora, 7, Local 20, E-18012 Granada, Spain
  \label{inst2}
  \and
  Centro de Astrobiolog\'ia (CSIC-INTA), Ctra. de Torrej\'on Ajalvir, km. 4, E-28850 Torrej\'on de Ardoz, Madrid, Spain
  \label{inst3}
  \and
  Joint ALMA Observatory, Av. El Golf 40, Piso 18, Las Condes, Santiago, Chile
  \label{inst4}
}

\abstract
{}
{
To derive carbon isotopic ratios from optically thin
tracers in the central regions of the starburst galaxies M~82 and NGC~253.
}
{
We present high sensitivity observations of CCH and two of its $^{13}$C isotopologues, C$^{13}$CH and $^{13}$CCH,
as well as the optically thin emission from
C$^{18}$O and $^{13}$C$^{18}$O.
We assume the column density ratio between isotopologues is representative of the $^{12}$C/$^{13}$C isotopic ratio.
}
{
From CCH, lower limits to the $^{12}$C/$^{13}$C isotopic ratio of 
138 in M~82, and 81 in NGC~253, are derived.
Lower limits to the $^{12}$C/$^{13}$C ratios from CO isotopologues support these.
$^{13}$C$^{18}$O is tentatively detected in NGC~253, which is the first reported detection in the extragalactic ISM. 
Based on these limits, we infer ratios of
$^{16}$O/$^{18}$O$>350$ and 
$>300$ in M~82 and NGC~253, respectively,
and $^{32}$S/$^{34}$S$>16$ in NGC~253.
The derived CCH fractional abundances toward these galaxies of
$\la1.1\times10^{-8}$ are in good agreement with those of
molecular clouds in the Galactic disk.
}
{
Our lower limits to the $^{12}$C/$^{13}$C ratio from CCH
are a factor of $2-3$ larger than previous limits.
The results are discussed in the context of molecular and nucleo-chemical evolution.
The large $^{12}$C/$^{13}$C isotopic ratio of the molecular ISM in these starburst galaxies suggest that the gas
has been recently accreted toward their nuclear regions.
}


\maketitle

\section{Introduction}
Isotopic ratios in the interstellar medium (ISM) of starburst (SB)
galaxies provide important clues on their nucleo-chemical evolution.
In particular, the $^{12}$C/$^{13}$C isotopic ratio
is believed
to be a good tracer of the chemical evolution of the Galaxy  (see e.g. Audouze 1985)
This ratio is understood to be a direct measurement of the primary to secondary nuclear processing \citep{Wilson1994}.
While $^{12}$C is a primary product of stellar nucleosynthesis, quickly produced via He burning in massive stars
that can be formed in first generation metal-poor stars, $^{13}$C is a secondary nuclear
product from $^{12}$C seeds \citep{Meyer1994,Wilson1992,Wilson1994}.

So far, only lower limits to the $\rm ^{12}C/^{13}C$ abundance ratios could be determined for some starburst galaxies.
A value of $\rm ^{12}C/^{13}C\ga40$ was derived toward the nuclei of the nearby galaxy NGC~253 based on
observations of CN, CS, and HNC \citep{Henkel1993} and further supported
by CO, HCN, and HCO$^+$ data on M~82 and NGC~4945 \citep{Henkel1993a,Henkel1994}. Additional CN data on M~82
and IC~342 resulted in lower limits to the ratio of $>40$ and $>30$ respectively \citep{Henkel1998}.
The CO and $^{13}$CO multi-transition non-LTE study toward M~82 by \citep{Mao2000} excludes $^{12}$C/$^{13}$C
ratios below 25 and point toward a ratio $>50$.
Apart from these multi-molecule studies toward the nearest SB galaxies, and limited by
sensitivity, only the $^{12}$CO/$^{13}$CO ratio is available for a small sample of nearby sources
\citep{Young1986,Sage1991a,Aalto1991,Casoli1992,Paglione2001}.
However, the lines from the species used so far can be severely affected by optically depth
effects as estimated by \citet{Henkel1993}.
Thus optically thin species are crucial to derive accurate isotopic ratios.

Ethynyl (CCH) has the brightest line in the 2~mm spectral scans carried out toward
NGC~253 and M~82 \citep[][Aladro et al. in prep.]{Mart'in2006a}.
Its hyperfine structure, unresolved
in these objects, can help to constrain the opacity through the observed lineshape.
Within the Galaxy, CCH emission appears tightly bound to massive star formation
at all evolutionary stages \citep{Beuther2008}.

In this paper, we present observations of CCH and two of its $^{13}$C isotopologues, and provide a more stringent lower limit to the
$^{12}$C/$^{13}$C isotopic ratio in the nuclear starburst ISM of M~82 and NGC~253.
Additionally we present CO observations in the $^{18}$O substitution and the double isotopologue
$^{13}$C$^{18}$O. The derived $^{12}$C/$^{13}$C ratio from CO, though not as stringent as that from CCH, supports the
idea of larger isotopic ratios in these starbursts than previously determined.

\section{Observations and Results}
\begin{figure}[!t]
\includegraphics[width=0.48\textwidth]{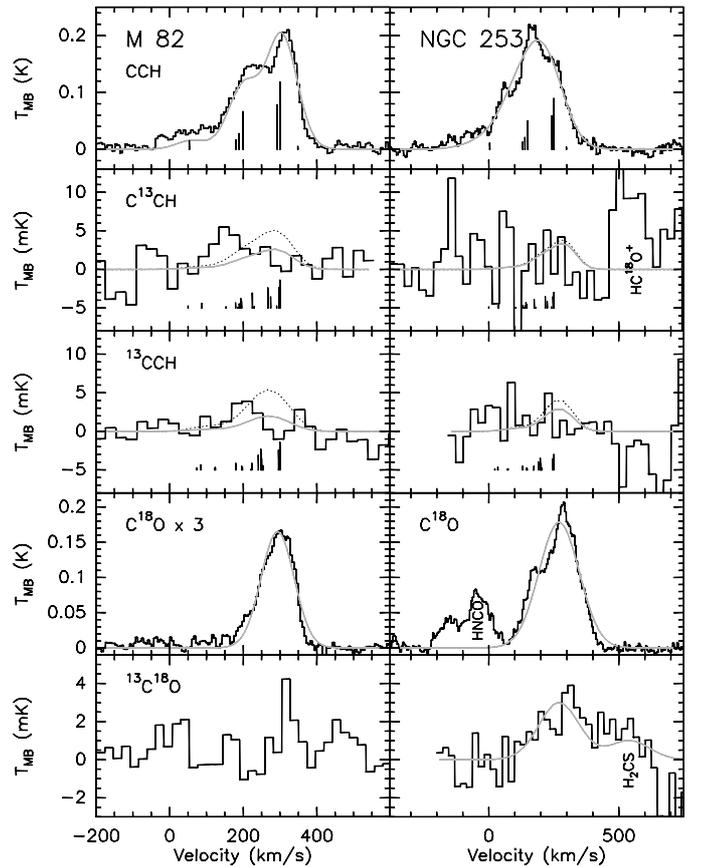}
\caption{
Observed spectra of the $J=2-1$ transitions of CCH, C$^{13}$CH, and $^{13}$CCH, and the $J=1-0$ transitions of 
C$^{18}$O and $^{13}$C$^{18}$O toward M~82 and NGC~253.
The intensity of the C$^{18}$O spectrum in M~82 has been multiplied by a factor of 3 for comparison and
visualization purposes.
The fit to the CCH and C$^{18}$O profiles are shown with grey lines.
For the undetected isotopologues we overlay the expected spectra assuming
a $^{12}$C/$^{13}$C isotopic ratio of 40 (dotted lines), as well as the upper limit profile derived in this work
(grey lines).
Vertical lines represent the relative intensities of the hyperfine components used to fit the detected CCH profiles,
and to calculate the upper limit to the intensities for the isotopologues.
$^{13}$C$^{18}$O is tentatively detected blended to a H$_2$CS feature toward NGC~253.
Spectra are presented at the original 4 and 2~MHz ($\sim 7\,\rm\,and\,5km\,s^{-1}$) resolution for CCH and C$^{18}$O, respectively.
C$^{13}$CH and $^{13}$CCH data were smoothed to 16~MHz ($\sim 28\,\rm km\,s^{-1}$) and $^{13}$C$^{18}$O to 8~MHz ($\sim 23\,\rm km\,s^{-1}$).}
\label{fig.CCHandCO}
\end{figure}

We have observed the $J=2-1$ group of hyperfine (hf) transitions of CCH (174.635~GHz),
C$^{13}$CH (170.491~GHz), and $^{13}$CCH (168.274~GHz), and the
$J=1-0$ transition of C$^{18}$O (109.782~GHz) and $^{13}$C$^{18}$O (104.711~GHz) toward M~82 and NGC~253.
Observations were carried out with the IRAM~30m telescope (Pico Veleta, Spain).
In the case of M~82, observations were aimed toward the north-eastern molecular lobe 
at the offset position $(+13'',+7.5'')$ from the center
($\alpha_{J2000}=09^{\rm h}55^{\rm m}51\fs9,\delta_{J2000}=69^\circ40'47\farcs1$).
NGC~253 observations were aimed at the central position ($\alpha_{J2000}=00^{\rm h}47^{\rm m}33\fs3,\delta_{J2000}=-25^\circ17'23\farcs15$)
for C$^{18}$O while the $^{13}$C$^{18}$O observations were slightly offset at $(+3'',-4'')$.
The beam widths at these frequencies were $\sim14''$ and $\sim23''$, for CCH and CO lines, respectively.
We used the wobbler switched observing mode with a symmetrical beam throw of $240''$ in azimuth and a wobbling frequency
of 0.5~Hz.
As spectrometers we used the $256\times4$~MHz filter banks and WILMA autocorrelator (2~MHz) for CCH and CO,
respectively.
The CCH data for NGC~253 were extracted from the 2~mm line survey by \citet{Mart'in2006a}.
Figs.~\ref{fig.CCHandCO} shows the observed spectra.

The hf structure of CCH is unresolved.
We fitted a comb of Gaussian profiles at the positions of the hf lines and intensity ratios fixed by their
spectroscopic parameters \citep{Muller2001}, shown as vertical lines in Fig.~\ref{fig.CCHandCO}.
In M~82, the wing of emission at low velocities in CCH is a contribution of both the fainter hf components
and the emission from the nuclear region, as also seen in the C$^{18}$O profile. 
Though two velocity components could be fit to the CCH profile in NGC~253 \citep{Mart'in2006a}, for the purposes
of this paper we just fit a single velocity component.
We assumed optically thin emission to fit the line profiles.
We show the resulting fits as grey lines in Fig~\ref{fig.CCHandCO} while the fitting parameters are shown in
Table~\ref{tab.parameters}.
Neither C$^{13}$CH nor $^{13}$CCH emission were detected toward M~82 and NGC~253.
No line contamination by other species is expected at the frequencies of the isotopologues.
We have estimated upper limits to the integrated intensities of the overall line profile of these isotopologues,
calculated from the corresponding spectral parameters \citep{Muller2001}.
We assumed a peak intensity of the overall profile at a $3\sigma$ level
(for 187 and 224~km\,s$^{-1}$ wide channels, corresponding to that from single Gaussian fit to the CCH features)
for M~82 and NGC~253, respectively.

Line emission from $^{13}$C$^{18}$O is not detected toward M~82 and only tentatively toward NGC~253.
This tentative $^{13}$C$^{18}$O detection, the first in the extragalactic ISM, appears blended to the 
H$_2$CS $3_{1,2}-2_{1,1}$ transition at 104.616~GHz. The derived intensity of $\sim1\,\rm mK$
supports 
the tentative detection reported by \citet{Mart'in2005}.

With the fitted integrated intensities we have estimated the beam averaged column densities for each species under
the local thermodynamic equilibrium assumption. An excitation temperature of $T_{\rm ex}=$10\,K was assumed for the
calculations.
The assumed temperature
, being the same for all isotopologues, has no effect on the derived column density ratios.
As reported from CO observations, the different isotopologues might be tracing different gas component
\citep{Wall1990,Aalto1994,Meier2000}, and therefore the similar excitation temperature assumption would not be fulfilled.
As a test case, if the $^{13}$C isotopologue was emitted from a gas component with $T_{\rm ex}\sim$50\,K, our assumption
would lead to the $^{12}$C/$^{13}$C ratio being overestimated by a factor of $\sim1.7$.
Multi transition observations of CCH and its $^{13}$C isotopologues would be needed to accurately establish the magnitude of such
effect. On the other hand, we do not expect it to be significant in the optically thin CO isotopologues.
The derived CCH and CO beam averaged column densities and upper limits are shown in Table~\ref{tab.parameters}.

The derived lower limits to the $^{12}$C/$^{13}$C ratios are of $>110$ and $>56$ for M~82 and NGC~253, respectively.
We can further constraint these limits by averaging the spectra of both C$^{13}$CH and $^{13}$CCH.
The resulting limit to the column densities raises the limits to the isotopic ratios up to
 $^{12}$C/$^{13}$C $>138$ and $>81$, respectively.
Using CO we obtain ratios of $>56$ and $\ga60$, respectively.

\begin{table*}
\caption{Observed lines parameters, and derived column densities and $^{12}$C/$^{13}$C ratios}
\label{tab.parameters}
\centering
\begin{tabular}{l l c c c c c c c c}
\hline
\hline
Source & Isotopologue             &   $J-J'$    & $\int{T_{\rm MB}{\rm d}v}$         & $v_{\rm LSR}$   & $\Delta v_{1/2}$\tablefootmark{a}& $T_{\rm MB}$\tablefootmark{a} &  $rms_{1\sigma}$\tablefootmark{b}     & $N$\tablefootmark{c}             & $\rm ^{12}C/^{13}C$ \\
       &                          &             & (K\,km\,s$^{-1}$)                  & (km\,s$^{-1}$)  &  (km\,s$^{-1}$)                  & (mK)                          &    (mK)                               & $(\times10^{13}\,\rm cm^{-2})$   &               \\
\hline                                                                                                                                                                                                                                                             %
M~82   & CCH                      &   $2-1$     & $34.2\pm0.6$                       &   307           &  97                              & 119                           &   7.6                                & $55\pm1$                         & ...           \\
       & C$^{13}$CH               &   $2-1$     & $<0.42$\tablefootmark{d}           & ...             & ...                              & ...                           &   2.3                                & $<0.7$                           & $>78$         \\
       & $^{13}$CCH               &   $2-1$     & $<0.30$\tablefootmark{d}           & ...             & ...                              & ...                           &   1.7                                & $<0.5$                           & $>110$        \\
       & $^{13}$CCH+C$^{13}$CH    &   $2-1$     & $<0.25$\tablefootmark{d}           & ...             & ...                              & ...                           &   1.5                                & $<0.4$                           & $>138$        \\
       & C$^{18}$O                &   $1-0$     & $6.2\pm0.1$                        &   293           & 106                              & 55                            &   1.9                                & $620\pm10$                       & ...           \\
       & $^{13}$C$^{18}$O         &   $1-0$     & $<0.11$                            &                 &                                  &                               &   0.8                                & $<11$                            & $>56$         \\
NGC~253& CCH                      &   $2-1$     & $45.8\pm0.6$                       & 216             & 171                              &  90                           &   8.4                                & $73\pm1$                         & ...           \\
       & C$^{13}$CH               &   $2-1$     & $<0.91$\tablefootmark{d}           & ...             & ...                              & ...                           &   4.5                                & $<1.6$                           & $>46$         \\                   
       & $^{13}$CCH               &   $2-1$     & $<0.73$\tablefootmark{d}           & ...             & ...                              & ...                           &   3.3                                & $<1.3$                           & $>56$         \\
       & $^{13}$CCH+C$^{13}$CH    &   $2-1$     & $<0.51$\tablefootmark{d}           & ...             & ...                              & ...                           &   2.2                                & $<0.9$                           & $>81$         \\
       & C$^{18}$O                &   $1-0$     & $34.7\pm0.2$                       & 270             & 183                              & 178                           &   3.7                                & $3470\pm20$                      & ...           \\
       & $^{13}$C$^{18}$O         &   $1-0$     & $\sim0.58\pm0.13$\tablefootmark{e} & 270             & 183                              &   3                           &   1.5                                & $58\pm13$                        & $\ga60$         \\
\hline                                                                                                                                                                                                                                   %
\end{tabular}
\tablefoot{
\tablefoottext{a}{This value refers to the brightest component in the group.}
\tablefoottext{b}{$1\sigma$ rms at the resolution shown in Fig.~\ref{fig.CCHandCO}.}
\tablefoottext{c}{Beam averaged column density assuming a $T_{\rm ex}=10$\,K.}
\tablefoottext{d}{$3\sigma$ limit to the integrated intensity} 
\tablefoottext{e}{Tentative detection blended to an H$_2$CS transition.
Velocity and width parameters were fixed to fit both lines simultaneously.}
}
\end{table*}
%
%

\section{$^{12}$C/$^{13}$C isotopic ratio from CCH and CO}

Estimating the interstellar isotopic ratio from the derived molecular abundances
one has to discuss
two main drawbacks, opacity and fractionation effects.

Lines of $^{12}$C isotopologues of CO and abundant species such as HCN or HCO$^+$, are likely
to be optically thick.
If our derived lower limits to the carbon isotopic ratio applies to CO, it would imply opacities of
$\tau_{\rm^{12}CO}\sim5$ and $\sim 9$ for NGC~253 and M~82, respectively,
as derived from the observed $^{12}$CO to $^{13}$CO integrated intensity ratios by \citet{Sage1991a}.
Higher opacities will lead to a decrease of the measured $^{12}$C/$^{13}$C ratio from CO and possibly
other abundant species.
Therefore, molecules with opaque lines
poorly constrain the $^{12}$C/$^{13}$C ratios.
From the hf fit to the line profile we can exclude large opacity effects affecting the CCH lines.
Moreover, small opacity effects on the CCH observed lines would result in an increase of our lower
limits to the $^{12}$C/$^{13}$C isotopic ratios.

Fractionation
\citep{Watson1976,Langer1984,Wilson1994} might produce an enhancement of $^{13}$CO in the outer layers
of molecular clouds. 
Thus the observed CO would be tracing regions with lower $^{12}$CO/$^{13}$CO ratio than the actual
$^{12}$C/$^{13}$C isotopic ratio.
The CCH observations reported by \citet{Sakai2010} toward the dark cloud TMC-1 and the star forming
region L1527, also show a $^{12}$C/$^{13}$C ratio a factor of $2-4$ larger than the interstellar
ratio. However, the chemical fractionation claimed in their work is mostly effective at very low
temperatures ($T_{\rm kin}\sim10$\,K) while isotopic exchange reactions rates tend to balance out at higher
temperatures \citep{Woods2009}. 
This fractionation is also expected to affect species other than CO, such as HCO$^+$ and HCN.
Moreover, the models by \citet{Woods2009} show how species such as HCN and HNC should result in
even larger ratios than those from CCH, which is not observed in galaxies.
Thus, chemical fractionation cannot explain the large CCH ratio observed in the warm
ISM in the central regions of galaxies. 
$^{12}$C/$^{13}$C isotopic ratios of  $20-25$ derived from CCH in the GC (Armijos et al. in prep.),
similar to those derived from other molecular species, support the hypothesis that fractionation
does not play an important role.

We conclude that ours limits for the $^{12}$C/$^{13}$C isotopic ratios derived from CCH isotopologues
are representative of the ISM in the observed SB galaxies.

\section{A large $^{12}$C/$^{13}$C isotopic ratio in starbursts}
Our results from CCH
show that the $^{12}$C/$^{13}$C isotopic ratio
in the SB environment is much higher than the value of $\sim20$ measured in the
Galactic Center (GC) region \citep{Wilson1994}.
Moreover, our limits are clearly higher than previous limits for
SB galaxies of $\sim40-50$ \citep{Henkel1993,Henkel1993a}.
We can exclude a ratio of 40 or lower, since in that case
we would have detected the $^{13}$CCH emission in both galaxies and that of C$^{13}$CH in M~82.
If the $^{12}$C/$^{13}$C ratios derived from CCH are representative of the overall $^{12}$C/$^{13}$C,
it would imply this ratio to be at least a factor of $2-3$ larger than previously
reported in the starburst ISM.
Though lower, the limits to the ratios we derive from CO isotopologues also suggest larger $^{12}$C/$^{13}$C
isotopic ratios than previously measured.

Single dish $^{12}$CO/$^{13}$CO ratios have been measured as high as 55 and
44 toward NGC~4195 and NGC~6240, respectively \citep{Casoli1992}.
However the latter was disproved by further
mapping resulting in a $^{12}$CO/$^{13}$CO ratio of $\sim 20$ \citep{Aalto2000}.
In Arp~299 (consisting of a merger of IC~694 and NGC~3690), a ratio of 60 was found toward the nucleus of
IC~694, and $\sim 10$ in the surrounding disk \citep{Aalto1999},
with the ratio ranging over $\sim20-35$ at single dish resolution.
Recently, a $^{12}$CO/$^{13}$CO $J=3-2$ ratio of $>40$ was reported toward the Cloverleaf quasar at $z\simeq2.6$ \citep{Henkel2010}


\section{Revisited carbon isotopic ratio implications}

\subsection{The oxygen $^{16}$O/$^{18}$O ratio}
The uncertainty in the measured $^{12}$C/$^{13}$C ratio directly affects the determination of the
$^{16}$O/$^{18}$O ratio based on measurements of $^{13}$CO and C$^{18}$O.
For NGC~253, \citet{Harrison1999} presented $^{13}$CO and C$^{18}$O measurements.
If we use our revised lower $^{12}$C/$^{13}$C limit of $>81$, their  $^{13}$CO/C$^{18}$O intensity ratio
would correspond to an abundance ratio of $^{16}$O/$^{18}$O$>300$
Toward M~82, using the $^{13}$CO observations from \citep{Mao2000}, observed at the same position, we derive a
$^{16}$O/$^{18}$O$>350$.
Both limits are now closer to the ratio of $\sim330$ found within the
4\,kpc molecular ring of the Milky Way \citep{Wilson1994}.

\subsection{The sulfur $^{32}$S/$^{34}$S ratio}
The $^{32}$S/$^{34}$S ratio based on measurements of $^{13}$CS
and C$^{34}$S will also be affected,
as that derived by \citet{Mart'in2005} for NGC~253.
Their value of the $^{32}$S/$^{34}$S ratio of $8\pm2$ would be doubled up to a value $>16$, closer to
the value of $\sim24$ measured in the Galactic disk \citep{Chin1996}.
The ratio of $^{32}$S/$^{34}$S$\sim13.5$ toward NGC~4945 \citep{Wang2004} might also be underestimated
as it was based on the CS/C$^{34}$S ratio assuming no significant saturation of the CS lines.

\subsection{H$_2$ column density determination}
Our large lower limits to the $^{12}$C/$^{13}$C ratio impacts the H$_2$ column density ($N_{H_2}$) derived
from the optically thin CO isotopologues.
These isotopologues, less affected by optical depth, are expected to be better suited to trace
the overall molecular material than the main isotopologue.
However, the determination of the $N_{H_2}$ will depend on the isotopic ratios.
If we apply our new limits to the oxygen isotopic ratio to estimate the $N_{H_2}$ from C$^{18}$O,
and assuming a conversion factor CO/H$_2\sim8.5\times10^{-5}$ \citep{Frerking1982},
we derive beam averaged limits of $N_{\rm H_2}>2.6\times10^{22}\,\rm cm^{-2}$ for M~82 and
$N_{\rm H_2}>1.2\times10^{23}\,\rm cm^{-2}$ for NGC~253.
We then infer CCH fractional abundances of $<2\times10^{-8}$ and $<0.6\times10^{-8}$, respectively,
in good agreement with the Galactic abundances found 
in prestellar cores \citep[$1\times10^{-8}$,][]{Padovani2009}, dark clouds \citep[$<6\times10^{-8}$,][]{Wootten1980},
diffuse clouds \citep[$3\times10^{-8}$,][]{Lucas2000}
and about two orders
of magnitude above the abundances in hot cores \citep[$2-8\times10^{-10}$,][]{Nummelin2000}.

Moreover, assuming our derived ratio of $^{16}$O/$^{18}$O$\sim300$
a better agreement is found between the gas mass estimate from 1.3~mm dust observations \citep{Krugel1990} 
and that from C$^{18}$O 
\citep[see Table~2 in][where a ratio of $^{16}$O/$^{18}$O$\sim150$ was used]{Mauersberger1996}.
This agreement further supports our result.

We can estimate a conversion factor of $X_{\rm CO}=N({\rm H_2})/\rm CO\sim 1.3\times10^{20}$\,cm$^{-2}$(K\,km\,s$^{-1}$),
using the CO $1-0$ integrated intensity of 920\,K\,km\,s$^{-1}$ \citep{Mauersberger1996}.
This conversion factor, significantly lower by a factor of $\sim 2$ than the ``standard'' value within the Galaxy of
$2-3\times10^{20}$\,cm$^{-2}$(K\,km\,s$^{-1}$) \citep{Solomon1987,Strong1988}, is a factor of 3 larger than the values proposed for the starburst
ISM of $0.3-0.4\times10^{20}$\,cm$^{-2}$(K\,km\,s$^{-1}$) \citep{Mauersberger1996,Mart'in2010}.
                                     
\section{Chemical evolution of starburst galaxies}

After several cycles of star formation, nucleosynthesis will enrich the interstellar medium with processed material
which could be ``dated'' by using the $^{12}$C/$^{13}$C isotopic ratio.
Thus, the $^{12}$C/$^{13}$C ratio is expected to decrease with time. 
The solar system ratio of 89 \citep{Wilson1994} is thought to be representative of the local ISM
when the Solar System formed, 4.6 billion years ago.
Although the $\rm^{12}CO/^{13}CO$ ratio toward both M~82 and NGC~253 is observed to be a factor of $\sim1.7$ larger
within the nuclear region than in the outer disk, this ratio is unlikely to be representative of the
$\rm^{12}C/^{13}C$ ratio due to opacity effects and/or variation of the CO/H$_2$ conversion factor, as reported by \citet{Paglione2001}.
On the contrary, a positive gradient towards the central region is expected, due to the molecular material
being further processed by stars, and similar to what is observed with optically thin tracers in the Milky Way \citep{Langer1990,Wilson1994}.
It is remarkable that, taking into account the ratios derived in this work,
the $^{12}$C/$^{13}$C ratio in SB galaxies is larger than in the Galactic center, 
the solar neighborhood \citep{Wilson1994} and even in the low metallicity ISM of the Large Magellanic Cloud
\citep[LMC; $^{12}$C/$^{13}$C$=49$,][]{Wang2009}.
Similarly, the oxygen and sulfur isotopic ratios measured do not evidence an enrichment of $^{18}$O and
$^{34}$S in the starburst ISM with respect to the values measured in the local ISM \citep{Wilson1994}.
Moreover, our lower limit to $^{16}$O/$^{18}$O is slightly above the value of 250 measured toward
the Galactic Center \citep{Wilson1994}, and the $^{32}$S/$^{34}$S would be at least a factor of 2 above
the expected ratio within the central 3\,kpc of the Galaxy from the gradient found by \citet{Chin1996}.
The $^{16}$O/$^{18}$O and $^{32}$S/$^{34}$S are significantly lower and similar, respectively, to those
measured in the LMC \citep[$^{16}$O/$^{18}$O=2000, $^{32}$S/$^{34}$S$\sim15$,][]{Wang2009}.

An enhanced of star formation was likely triggered across the disk of M~82 as a result of a close
encounter with M~81 about 220~Myr ago. However, the ongoing nuclear starbursts started $\sim15$\,Myr ago, with
nuclear star clusters dated between 7 and 15~Myr \citep{Konstantopoulos2009}.
This dating is in agreement with the two short duration bursts of star formation in the nuclear region of M~82,
with peaks at 5 and 10~Myr \citep{ForsterSchreiber2003}.
The nucleus of NGC~253 appear to host an even younger starburst with a median age of $\sim6$\,Myr
\citep{Fern'andez-Ontiveros2009}.
With a average lifetime of 10~Myr, massive star clusters formed in the starburst appear not to have efficiently
enriched the ISM in their nuclear region.
Furthermore, the enriched material could have been banished from the nuclear region via the starburst driven super-winds in both
galaxies \citep{Heckman1990}.

One possible interpretation would be that in very young SBs one should find 
large $^{12}$C/$^{13}$C ratios due to fast evolution of very massive stars,
leading to a overproduction of $^{12}$C relative to $^{13}$C, synthesized in intermediate mass stars
\citep{Henkel1993a}.
However, such fast evolution might also result in an overproduction of $^{18}$O \citep{Henkel1993a}
which is not observed.
Moreover, if the time scale of the main burst in the nucleus of M~82 is less than 15~Myr
\citep{ForsterSchreiber2003,Konstantopoulos2009}, then most of the $^{12}$C enriched 
material would be in the hot outflow with contribution to the molecular gas.
The time scale is rather short to have the $^{12}$C enriched material
well mixed over the $\sim$300~pc of our beam size, in both M~82 and NGC~253.
It would mean that the enhanced $^{12}$C molecular gas is confined to shells around the stellar 
clusters with a huge $^{12}$C/$^{13}$C ratio.

Thus, our measured isotopic ratios suggest that the bulk of the mass of molecular material within
the nuclear regions of M~82 and NGC~253 consists of little or unprocessed gas by massive stars.
This suggests that the ISM which is supporting the starbursts in M~82 and NGC~253 has been recently funneled toward their
centers from the outer regions or accreted by these galaxies.
The evolution of these isotopic ratios as a function of metallicity is not clear from the available observations.
However, \citet{Wang2009} found the $^{16}$O/$^{18}$O ratio to be a good tracer of metallicity.
The lower limits to the ratio found toward M~82 and NGC~253, larger than the value toward the Galactic Center,
is consistent with that found within the 4\,kpc molecular ring and even that within the local ISM.
This would support the idea of inflowing material from the outer regions of these galaxies toward their centers
conforming the gas concentration unleashing their starburst events,
where enriched material from previous star formation events would contribute little to the bulk of the 
starburst molecular material.


\begin{acknowledgements}
This work has been partially supported by the Spanish Ministerio de Ciencia e Innovaci\'on
under projects ESP2007-65812-CO2-01 and AYA2008-06181-C02-02.
\end{acknowledgements}

\bibliographystyle{aa}	
\bibliography{1213CCH.bib}	

\end{document}